\documentclass[useAMS,usenatbib]{mn2e}

\usepackage{txfonts,graphicx,amssymb,rotating}

\title[The missing gas disc at the Galactic Centre]{Galactic Centre star formation: the case of the missing gas disc}

\author[R.D.Alexander et al.]
{R.D.Alexander$\thanks{email: richard.alexander@leicester.ac.uk}$, S.L.Smedley, S.Nayakshin \& A.R.King
\\Department of Physics \& Astronomy, University of Leicester, Leicester, LE1 7RH}

\begin{document}
\voffset=-0.5in
\newcommand{\Msunyr}{M$_{\odot}$yr$^{-1}$}
\newcommand{\Msun}{M$_{\odot}$}

\pagerange{\pageref{firstpage}--\pageref{lastpage}} \pubyear{2011}

\date{Accepted 2011 September 18.  Received 2011 September 16; in original form 2011 August 19}

\maketitle

\label{firstpage}

\begin{abstract}
We study the dynamical evolution of stars and gas close to the centre of the Milky Way.  Any plausible means of forming the young stars observed at the Galactic Centre leaves behind a residual gas disc at $\sim 0.01$pc radii.  We show that the combined effects of viscous accretion and gravitational interactions with stars do not remove the residual gas efficiently, and that a substantial gas disc, interior to the stellar disc, persists for $>10$Myr after the stars form.  Since no such disc is currently seen at the Galactic Centre we argue that it has been accreted by the super-massive black hole.   This scenario offers an attractive connection between nuclear star formation and black hole feeding, and we suggest that the `missing' gas may have been used to power Sgr A$^*$.  
\end{abstract}



\section{Introduction}\label{sec:intro}
The discovery of a large cluster of young stars within 0.1pc of Sgr A$^*$, the super-massive black hole (SMBH) at the centre of the Galaxy, poses a number of interesting questions.  The physical conditions so close to the Galactic Centre (GC) are extreme, and the high temperatures and tidal shear prohibit the formation of stars by the same mechanisms that occur in the solar neighbourhood.  However, observations show that tens of B-type stars orbit within $\simeq 0.01$pc of the SMBH, while a larger population of $\sim$100 O- and Wolf-Rayet (WR) stars exists at slightly larger radii, $\simeq 0.1$pc  \citep[e.g.,][]{genzel03,ghez05}.  These stars are young ($\lesssim 10$Myr for the O- and WR stars), which suggests that they must have formed at or close to their current location.  How stars form in such an extreme environment remains an open question.

An important additional clue to the origin of the young stars at the GC comes from their dynamics.  Many of the O- and WR stars are known to orbit Sgr A$^*$ in a single coherent disc \citep[e.g.,][]{paumard06,lu09,gillessen09}, and this provides observational support for the idea that stars can form in accretion discs around SMBHs \citep[e.g.,][]{pac78,ks80,goodman03}.  Such discs are expected to be gravitationally unstable at large radii, and in recent years a number of authors have suggested that disc fragmentation due to gravitational instability is responsible for the young stellar disc at the GC \citep[e.g.,][]{lb03,nayak06,levin07}.  It is also tempting to link GC star formation to the broader question of SMBH accretion and growth \citep{kp06,kp07}, but the relationship between nuclear star formation, accretion and black hole feeding remains poorly understood.  

The disc fragmentation scenario works as follows.  Some sort of `accretion event', such as the capture of a molecular cloud \citep[e.g.,][]{wyz08}, results in gas with very low angular momentum falling towards the SMBH.  This gas dissipates energy through shocks and radiative cooling, and forms an accretion disc around the SMBH.  The gas at small radii accretes on to the SMBH, but beyond some critical radius the disc is gravitationally unstable and fragments to form stars.  Numerical simulations \citep[e.g.,][]{nayak07,rda08,br08,hn09} have shown that this scenario is broadly consistent with the observed properties of the GC stellar disc, but how the disc-star system evolves after its formation is far from certain.  

The fate of the residual gas disc is particularly interesting.  Analytic arguments \citep[e.g.,][]{nc05,kp07} show that in the case of Sgr A$^*$ the accretion disc fragments into stars only at radii $R \gtrsim 0.03$--0.05pc, which is coincident with the inner edge of the observed stellar disc \citep[e.g.,][]{paumard06,lu09}.  Inside this radius the gas disc is gravitationally stable, and does not fragment into stars.  Numerical simulations of gas deposition into the inner parsec find that a significant mass of gas, $\sim 10^3$--$10^4$\Msun, is deposited into this inner region \citep[e.g.,][]{br08,hn09}, and it has been argued that the residual gas disc could have been present for a long time, perhaps until the current epoch, supplying Sgr A$^*$ with gas at moderate rates \citep{nc05,hn09}.  However, no such gas disc is observed today.  An optically thick accretion disc at $\simeq 0.01$pc radii is completely ruled out by observations \citep[see, e.g.,][]{cuadra03,nayak04}, and the low luminosity of Sgr A$^*$ itself also sets very strict limits on the rate of gas accretion in any extant disc \citep[e.g.,][]{bower03}.  The fate of this `missing' gas disc remains an open question, and may provide an important link between nuclear star formation and SMBH accretion and growth.

In this paper we model the evolution of this residual gas disc.  Our model is presented in Section \ref{sec:model}.  The model includes viscous accretion of gas and tidal torques from the stellar disc, and the stars interact gravitationally with one another.  Our results, presented in Section \ref{sec:results}, show that these processes do not operate rapidly enough to remove the residual gas disc in $\sim10$Myr.   We compare our results to observations and discuss possible explanations for the missing gas in Section \ref{sec:discussion}, and summarize our conclusions in Section \ref{sec:summary}.


\section{Model}\label{sec:model}
Our model consists of a thin accretion disc around a SMBH, interior to a co-planar disc of stars.  The disc accretes due to viscous transport of angular momentum, and the stars exchange angular momentum with the disc through the action of tidal torques.  We further include the interaction of the stars with one another, which is known to modify their dynamics significantly over Myr periods \citep{rda07}.  Our initial conditions are consistent with formation of the stars via disc fragmentation, and we use a one-dimensional numerical scheme to follow the non-linear evolution of this system forward in time.

\subsection{Disc model}
The evolution of the gas surface density, $\Sigma(R,t)$ in the accretion disc is governed by the equation \citep{lbp74,pringle81}
\begin{equation}\label{eq:diff}\label{eq:gas_evo}
\frac{\partial \Sigma}{\partial t} = \frac{1}{R}\frac{\partial}{\partial R}\left[ 3R^{1/2} \frac{\partial}{\partial R}\left(\nu \Sigma R^{1/2}\right) - \sum_{i=1}^{N_*} \frac{2 \Lambda_i \Sigma R^{3/2}}{(GM_{\mathrm {bh}})^{1/2}}\right] .
\end{equation}
Here $t$ is time, $R$ is cylindrical radius, $\nu$ is the kinematic viscosity and $M_{\mathrm {bh}} = 4\times10^6$\Msun~is the mass of the central SMBH.  The first term on the right-hand side represents viscous evolution of the disc, while the second term describes the response of the disc to the tidal torque from the stars.  $\Lambda_i(R,a_i)$  is the rate of specific angular momentum transfer from the $i$th star to the disc, and the total torque on the disc is found by summing over all $N_*$ stars.  The torque from a single star on the disc is essentially the same as that found in Type II planet migration and, following \citet{trilling98} and \citet{armitage02}, for a star of mass $M_{*,i} = q_i M_{\mathrm {bh}}$ at radius (semi-major axis) $a_i$ we adopt the following form for $\Lambda_i$
\begin{equation}\label{eq:lambda}
\Lambda_i(R,a_i) = \left\{ \begin{array}{ll}
- \frac{q_i^2 GM_*}{2R} \left(\frac{R}{\Delta_{\mathrm p,i}}\right)^4 & \textrm{if } \, R < a_i\\
\frac{q_i^2 GM_*}{2R} \left(\frac{a}{\Delta_{\mathrm p,i}}\right)^4 & \textrm{if } \,R > a_i\\
\end{array}\right.
\end{equation}
Here
\begin{equation}
\Delta_{\mathrm p,i} = \textrm{max}(H,|R-a_i|) \, ,
\end{equation}
and $H = c_{\mathrm s}/\Omega$ is the disc scale-height.  This form for $\Lambda_i$ is the same as that used by \citet[][see also \citealt{lp79}]{lp86}, but modified to give a symmetric treatment inside and outside the star's orbit (though in practice only the $R<a_i$ case is relevant here).  This transfer of angular momentum causes the stars to migrate at a rate
\begin{equation}\label{eq:da_dt}
\frac{da_i}{dt} = - \left(\frac{a_i}{GM_{\mathrm {bh}}}\right)^{1/2} \left(\frac{4\pi}{M_{*,i}}\right) \int R\Lambda_i \Sigma dR \, .
\end{equation}
We adopt a standard \citet{ss73} $\alpha$-viscosity 
\begin{equation}
\nu = \alpha  c_{\mathrm s} H \, ,
\end{equation}
where $c_{\mathrm s}$ is the local sound speed, $\Omega = \sqrt{G M_{\mathrm {bh}}/R^3}$ is the Keplerian angular frequency, and the dimensionless parameter $\alpha$ specifies the efficiency of angular momentum transport \citep[thought to be due to the magnetorotational instability, e.g.,][]{bh91,bh98}.  We initially adopt a constant value $\alpha = 0.1$, the fiducial value for black hole discs \citep*[e.g.,][]{klp07}.

For the disc's thermal structure we use a simplified form for the energy equation \citep[e.g.,][]{cannizzo93}
\begin{equation}\label{eq:temp}
\frac{\partial T_{\mathrm c}}{\partial t} = \frac{2 (Q_+ - Q_-)}{c_{\mathrm P}\Sigma} - u_R
\frac{\partial T_{\mathrm c}}{\partial R} \, .
\end{equation}
Here $T_{\mathrm c}$ is the midplane temperature and $Q_+$ and $Q_-$ are the instantaneous heating and cooling rates.  The radial velocity $u_R$ in the advective term in Equation \ref{eq:temp} is assumed to be the vertically averaged value
\begin{equation}\label{eq:u_r}
u_R = - \frac{3}{\Sigma R^{1/2}} \frac{\partial}{\partial R} \left(\nu \Sigma R^{1/2}\right) \, .
\end{equation}
Here $c_{\mathrm P}$ is the specific heat capacity: we adopt the functional form for $c_{\mathrm P}(T_{\mathrm c})$ from \citet{cannizzo93}, but note that the heat capacity is approximately constant for the low temperatures considered here.  The rate of heating due to viscous disspation, $Q_+$ is given by
\begin{equation}
Q_+ = \frac{9}{8}\nu \Sigma \Omega^2 \, .
\end{equation} 
We adopt a `one-zone' model for cooling \citep[as used by, for example,][]{jg03}, where
\begin{equation}\label{eq:onezone}
Q_- = \frac{8}{3} \frac{\tau}{1 + \tau^2} \sigma_{\mathrm {SB}} \left(T_{\mathrm c}^4 - T_{\mathrm {min}}^4\right) \, .
\end{equation}
Here, the second term allows for a smooth transition between the limits of optically thick and optically thin cooling, $\sigma_{\mathrm {SB}}$ is the Stefan-Boltzmann constant, and the last term is prescribed in a manner than enforces a minimum disc temperature.  Due to the strong local radiation field from the stellar cluster we expect a high background temperature close to Sgr A$^*$, so we adopt $T_{\mathrm {min}} = 50$K throughout \citep[e.g.,][]{levin07}.  The vertical optical depth $\tau$ is computed as
\begin{equation}
\tau = \frac{1}{2}\kappa(T_{\mathrm c},\rho_{\mathrm c}) \Sigma \, ,
\end{equation}
with the midplane density $\rho_c$ evaluated from the surface density by assuming that the disc is vertically isothermal:
\begin{equation}
\rho_{\mathrm c} = \frac{\Sigma}{\sqrt{2\pi} H} \, .
\end{equation}
The local sound speed is essentially a means of keeping track of the local pressure, and includes contributions from both gas and radiation pressure:
\begin{equation}\label{eq:c_s}
c_s^2 = \frac{{\cal R} T_c}{\mu} + \frac{4 \sigma_{{\mathrm {SB}}} T_c^4}{3 c \rho_c}
\end{equation}
Here $\cal R$ is the gas constant, $\mu$ is the mean molecular weight, and $c$ is the speed of light.  In the region of interest ($R \sim 0.01$pc) our disc remains cold, with $T_c \lesssim 1000$K, and radiation pressure is negligible.  As a result we thus fix $\mu = 2.3$, consistent with a disc of molecular gas.  At these temperatures and densities the opacity is dominated by dust, and we use the numerical fits for $\kappa(T,\rho)$ given by \citet{zhu07,zhu08}.

\subsection{Stellar scattering}\label{sec:scattering}
In addition to feeling tidal torques from the disc, the stars also interact gravitationally with one another.  Stellar interactions act to increase the stellar velocity dispersion, which has the dual effect of thickening the stellar disc and increasing the eccentricities of the individual stellar orbits.  In practice this process is three-dimensional, but previous studies have shown that as long as the disc is thick enough to fall into the `dispersion-dominated' regime then a simple analytic approach is sufficiently accurate to model the stellar scattering process \citep[e.g.,][]{rda07}.  For a population of stars with a single mass $M_*$ in a disc around a black hole, the stellar velocity dispersion $\sigma_*$ evolves as
\begin{equation}\label{eq:vel_disp}
\frac{d\sigma_*}{dt} = \frac{\sigma_*}{t_{\mathrm r}} = \frac{G^2 N_*^2 M_*^2 \ln \Pi_*}{C \bar{R} \Delta R t_{\mathrm {orb}} \sigma_*^3} \, .
\end{equation}
Here $t_{\mathrm r}$ is the (two-body) relaxation time-scale, $\bar{R}$ is the mean stellar radius (semi-major axis), $\Delta R$ is the radial extent of the stellar disc (computed as the standard deviation of $a_i$), and $t_{\mathrm {orb}}$ is the Keplerian orbital period at $\bar{R}$.  $\Pi_* = \sigma_*^2 \Delta R/2GM_*$ is the usual Coulomb logarithm\footnote{In order to avoid confusion we use $\Pi_*$ for the Coulomb logarithm, as the more usual symbol ($\Lambda$) represents the tidal torque function here.}, and $C$ is a numerical scaling constant which \cite{rda07} found to be $C \simeq 2.2$.

In the absence of any gas two-body scattering causes the velocity dispersion to increase roughly as $\sigma_* \propto t^{1/4}$, but in the presence of a gas disc the effects of stellar scattering act in competition with the tidal torques described in Equations \ref{eq:gas_evo}--\ref{eq:da_dt}.  We approximate the effect of stellar scattering in our one-dimensional disc-star code by applying stochastic `kicks' in radius to the individual stars.  Based on the relationship between velocity dispersion and eccentricity, we define a kick size
\begin{equation}
\Delta R_{\mathrm k} = \frac{1}{\sqrt{2}} \bar{R} \frac{\sigma_*^2}{v_{\mathrm K}^2} \, ,
\end{equation}
where $v_{\mathrm K} = \sqrt{GM_{\mathrm {bh}}/\bar{R}}$ is the Keplerian orbital speed at radius $\bar{R}$.  The stars are kicked $N_{\mathrm k}$ times per relaxation time $t_{\mathrm r}$, with the individual kick sizes drawn randomly from the range
\begin{equation}
-\frac{\Delta R_{\mathrm k}}{\sqrt{N_{\mathrm k}}} \le \delta a_{i,\mathrm k} \le \frac{\Delta R_{\mathrm k}}{\sqrt{N_{\mathrm k}}} \, .
\end{equation}
This form is chosen so that rate of increase of the velocity dispersion is independent of the kick frequency.  The value of $N_{\mathrm k}$ therefore controls only the level of discreteness of the numerical integration.  Test calculations indicate that the value of $N_{\mathrm k}$ does not strongly influence our results, and for computational convenience we choose $N_{\mathrm k} = 1000$.

\subsection{Initial conditions}\label{sec:ICs}
We choose our initial conditions to reflect the conditions immediately after a gravitationally unstable gas disc has fragmented into stars.  Any plausible accretion disc model has a \citet{toomre64} $Q$ parameter which decreases with increasing radius \citep[e.g.,][]{nayak06,levin07,kp07}, and consequently disc fragmentation only happens beyond some critical radius.  We choose our initial disc surface density profile to be a power-law, truncated at some outer radius $R_{\mathrm {out}}$:
\begin{equation}
\Sigma(R) \propto R^{-p} \qquad , \qquad R \le R_{\mathrm {out}}
\end{equation}
and zero at larger radii.  We then randomly assign the initial stellar radii $a_i$ in the range $[R_{\mathrm {out}}, 2R_{\mathrm {out}}]$. In order to prevent Equation \ref{eq:vel_disp} from diverging we assign a small initial velocity dispersion $\sigma_*(0) = 20$km/s [though our results are not at all sensitive to the exact value of $\sigma_*(0)$].

Based on observations of the stellar discs at the GC, we model $N_* = 100$ stars of $M_{*,i} = 100$\Msun.  If these stars formed via disc fragmentation, the outer edge of the accretion disc should be located at the outermost point where the disc remains gravitationally stable.  We therefore normalise the disc mass by demanding that the disc be marginally gravitationally stable at its outer radius, and set $Q=2$ at $R = R_{\mathrm {out}}$.  In thermal equilibrium the disc temperature depends only on the orbital frequency $\Omega$ and the optical depth $\kappa$, so fixing the value of $Q$ in this manner uniquely determines the initial surface density $\Sigma(R_{\mathrm {out}})$.  Our model therefore has only two free parameters: the outer disc radius $R_{\mathrm {out}}$, beyond which the disc is assumed to have fragmented into stars, and the power-law index $p$.  Additional models were also computed in which the initial condition was a gas `ring', truncated inside some inner radius $R_{\mathrm {in}}$ (see Section \ref{sec:spikes}).

\subsection{Numerical Method}
We solve Equations \ref{eq:diff} and \ref{eq:temp} using an explicit finite-difference method, on a grid that is equispaced in $R^{1/2}$ \citep[e.g.,][]{pvw86}, using 3437 grid cells spanning the range $[0.003\mathrm {pc},1.0\mathrm {pc}]$.  We use a staggered grid, with scalar quantities evaluated at zone centres and vectors (velocities) at zone faces.  The derivative in the advective term (Equation \ref{eq:u_r}) is evaluated as the first-order, up-winded value.  The time-step is typically limited by the radial velocity of the gas very close to the location of the stars, and evolving the system on such a short time-step is unnecessary if the gas surface density here is very low.  Consequently we impose a maximum torque per star (and therefore a maximum gas velocity in the radial direction) of $|\Lambda_i| \le 0.1 R H \Omega^2$.  We adopt zero-torque boundary conditions at both the inner and outer boundaries, by setting $\Sigma=0$ in the boundary cells.  Equation \ref{eq:vel_disp} is also integrated explicitly with the same time-step and, as discussed above, the stellar velocity dispersion is modelled by giving the stars random `kicks' $N_{\mathrm k} = 1000$ times per relaxation time.

\section{Results}\label{sec:results}
\subsection{Disc models}\label{sec:discs}
We computed a grid of models spanning the range of plausible values for the two free parameters $R_{\mathrm {out}}$ and $p$.  The models used $p = 1/2$, 1, 3/2, 2, and $R_{\mathrm {out}} = 0.02$, 0.04pc, making a total of 8 individual runs.  Each of which was evolved up to a time $t=10$Myr, which is approximately the age of the observed GC stellar disc(s).

\begin{figure}
\centering
       \resizebox{\hsize}{!}{
       \includegraphics[angle=270]{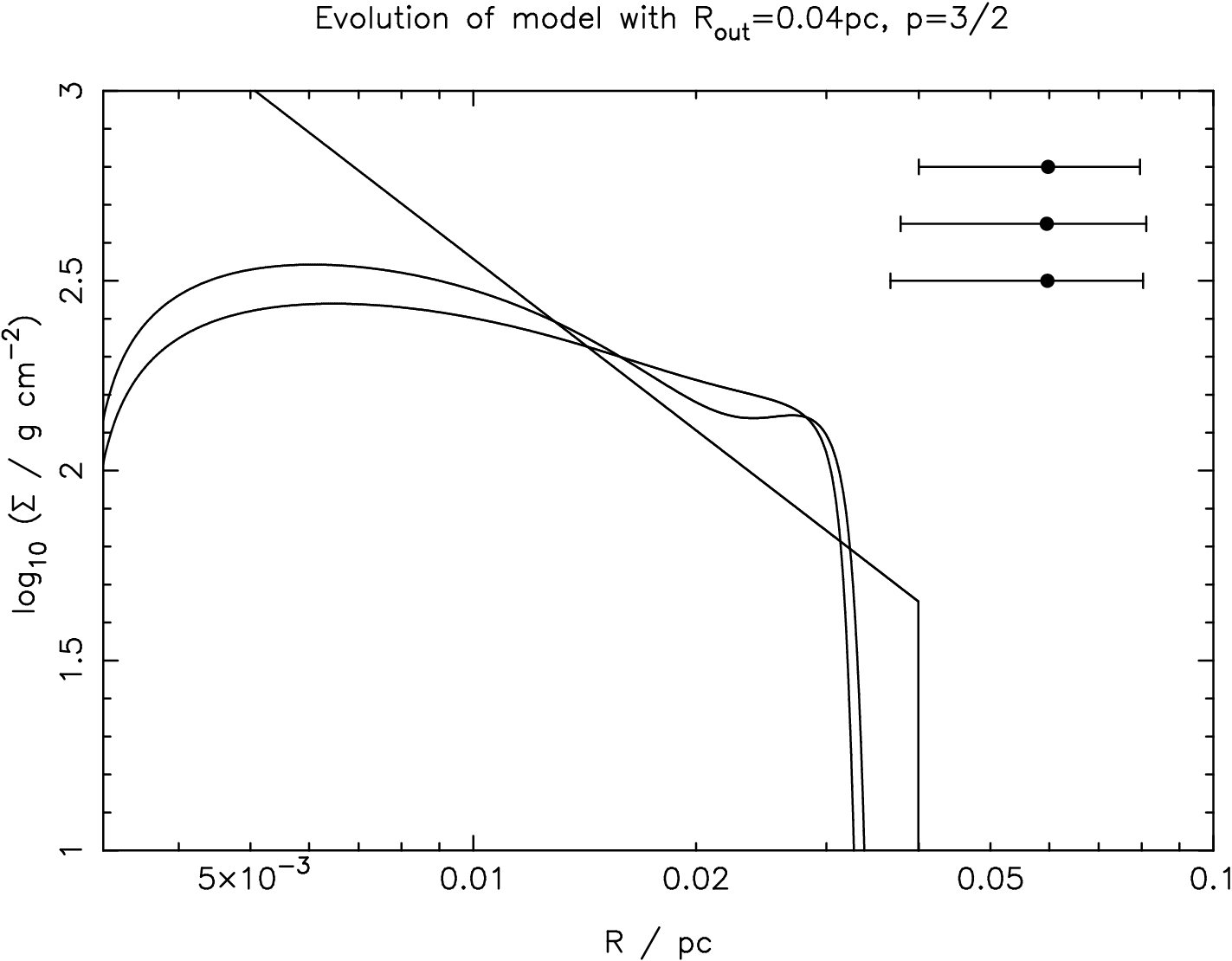}
       }
       
       \vspace*{6pt}
       
       \resizebox{\hsize}{!}{
       \includegraphics[angle=270]{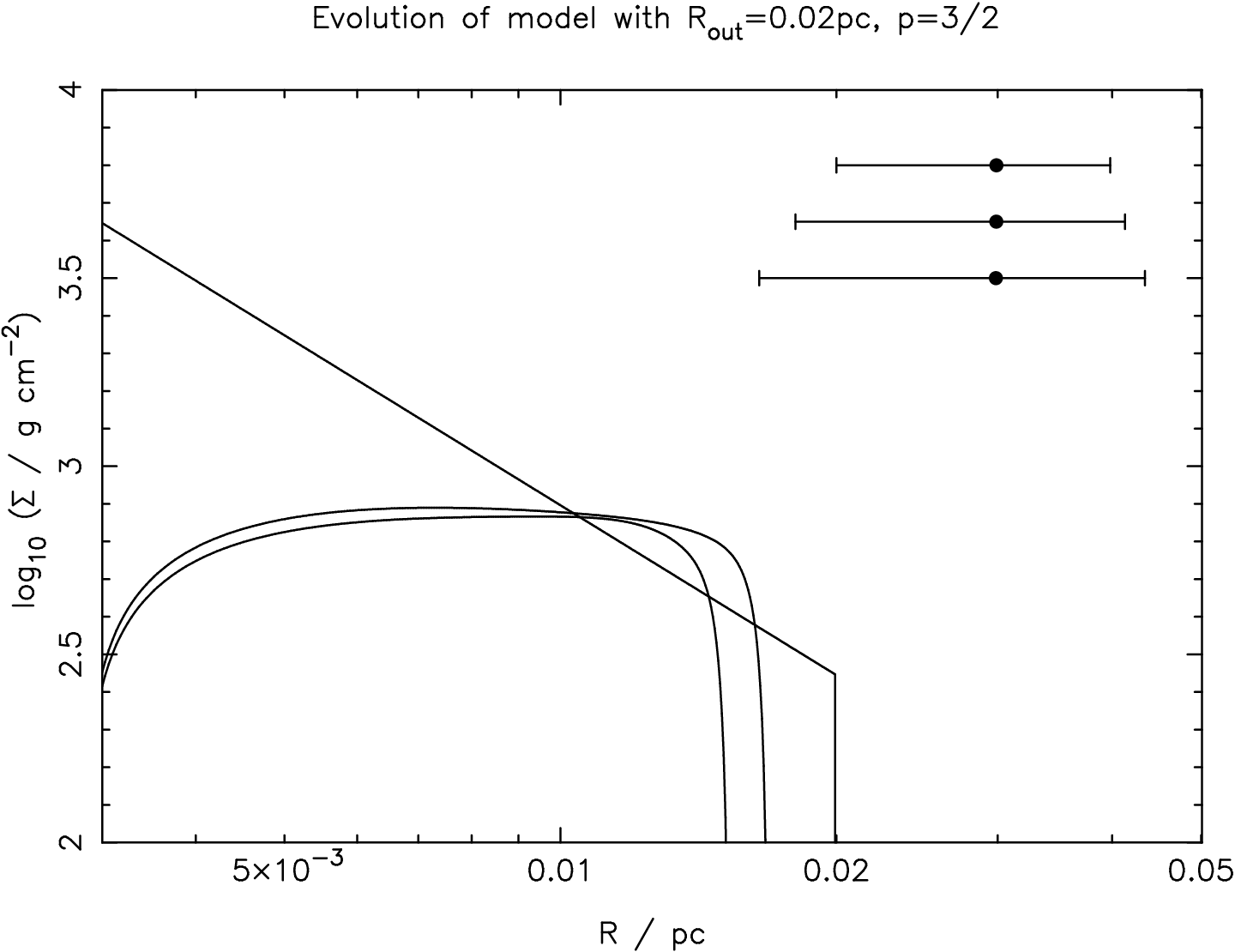}
       }
       \caption{Evolution of the disc-stars system: the upper panel shows the model with $R_{\mathrm {out}} = 0.04$pc, $p=3/2$; the lower panel the model with $R_{\mathrm {out}} = 0.02$pc, $p=3/2$.  In each case the lines show the disc surface density, plotted at $t=0$, 5 \& 10Myr.  The filled circles show the mean orbital radius of the stars, with the `error bar' denoting the extent of the stellar disc (extending from the minimum to maximum stellar orbital radii): the top, middle and bottom points again represent $t=0$, 5 \& 10Myr.  Stellar scattering increases the width of the stellar ring, while tidal and viscous torques drive accretion in the gas disc.  However, in both cases the system only undergoes modest evolution: 10Myr after formation the stars have not migrated significantly from their initial positions, and a substantial gas disc remains.  The decrease in surface density at very small radii is an artefact of our (zero-torque) inner boundary condition, and is not significant.}     
           \label{fig:sigma}
\end{figure}

Fig.\ref{fig:sigma} shows the evolution of the models with $p=3/2$.  The presence of the stars provides a strong tidal barrier, and the torques from the stars prevent viscous spreading of the gas disc to radii $>R_{\mathrm {out}}$.  The back reaction of the disc on the stars is small (because the torque function declines very sharply with increasing star-disc separation; see Equation \ref{eq:lambda}), so only the inner part of the stellar disc is strongly affected by the gas.  For the majority of the stars stellar scattering dominates over the star-disc torques, so the stellar discs spread in radius.  The model with $R_{\mathrm {out}}=0.02$pc evolves more quickly than the model with $R_{\mathrm {out}}=0.04$pc, because the dynamical and viscous time-scales are shorter at smaller radii.  However, in both cases the systems undergo only modest evolution over the 10Myr duration of the calculation: the stellar discs spread in radius by $<25$\%, and a significant gas disc remains out to radii $>0.01$pc.

These results are not substantially altered when we change the surface density slope $p$.  Discs with lower values of $p$ have less mass at small radii and generally evolve more quickly, but the overall trends seen in Fig.\ref{fig:sigma} hold for all of our models.  On time-scales $\sim$10Myr the stellar discs do not migrate significantly from their initial configuration, and the combined effect of viscous accretion and tidal torques has only a modest effect on the residual gas disc.  All of our models predict that a substantial gas disc extends to large radii for $>10$Myr, with surface densities $\Sigma > 100$g\,cm$^{-2}$ at $R \sim 0.01$pc.  These surface densities correspond to large optical depths ($\tau > 100$), and all of our discs are extremely optically thick at $\sim 0.01$pc.  By contrast, the presence of an optically thick gas disc at these radii is strongly ruled out by observations of the GC \citep{cuadra03,paumard04}, so at first sight our models are not consistent with the data.

\begin{figure}
\centering
       \resizebox{\hsize}{!}{
       \includegraphics[angle=270]{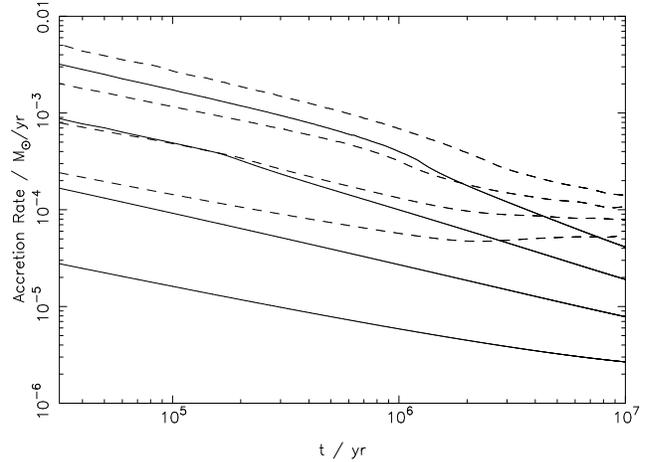}
       }
              \caption{Accretion rate as a function of time.  Solid lines represent the models with $R_{\mathrm {out}}=0.04$pc; dashed lines $R_{\mathrm {out}}=0.02$pc.  From bottom to top, in both cases, the curves show the models with $p=1/2$, 1, 3/2 \& 2, respectively.  In all cases substantial accretion on to the SMBH persists for $>10$Myr after the stars form.  For reference, the Eddington accretion rate is $\dot{\mathrm M}_{\mathrm {Edd}}\simeq 0.08$\Msunyr.}     
           \label{fig:mdot}
\end{figure}

Fig.\ref{fig:mdot} shows the accretion rate on to the central SMBH as a function of time in the different models.  We see the power-law behaviour characteristic of viscous accretion discs \citep[e.g.,][]{lbp74}, modified somewhat by the stellar tidal torques.  In all cases significant accretion on to the SMBH persists for tens of Myr.  For reference, if we assume an accretion efficiency of $\eta = 0.1$ the Eddington rate for a $4\times10^6$\Msun~SMBH is $\dot{\mathrm M}_{\mathrm {Edd}}\simeq 0.08$\Msunyr.  The initial accretion rates exceed 0.01$\dot{\mathrm M}_{\mathrm {Edd}}$ at early times, and even the lowest mass discs are still accreting at $\sim 10^{-3}$--$10^{-4}\dot{\mathrm M}_{\mathrm {Edd}}$ 10Myr after the outer disc fragmented into stars.  This is in stark contrast with the current state of Sgr A$^*$, where the accretion rate is measured to be $< 10^{-6}\dot{\mathrm M}_{\mathrm {Edd}}$ \citep{bower03}.  Both the gas surface densities and SMBH accretion rates show our discs simply do not accrete quickly enough to satisfy the observations, and we conclude that some other mechanism must remove the residual gas left over from the formation of the GC stellar disc.

\subsection{Additional models}\label{sec:spikes}
The most obvious solution to the problem of too much residual gas is that the initial configuration was not a disc, with gas extending all the way to small radii, but rather a narrow ring close to the radius where the stars formed.  This configuration is the likely outcome if the disc and stars formed from an initial gas cloud with a narrow dispersion in its angular momentum, and allows us to place a more interesting limit on the post-star-formation conditions at the GC.

In order to test this hypothesis we ran additional models where the initial conditions for the gas disc were a narrow `spike' in the gas surface density just interior to the stellar disc.  The set-up was as before, but now with the surface density set to zero both beyond $R_{\mathrm {out}}$ and also inside $R_{\mathrm {in}} = 0.9R_{\mathrm {out}}$.  The surface density was assumed to be constant (i.e., $p=0$), but we note that with such a narrow dynamic range in radius the surface density slope has a negligible effect on these models.  As before we ran models with $R_{\mathrm {out}} = 0.02$, 0.04pc, and this time set the viscosity parameter $\alpha = 1$.  This choice of $\alpha$ is deliberately high, and serves as an upper limit to the efficiency of viscous accretion.  The total initial gas masses in these models are 365\Msun~(for $R_{\mathrm {out}} = 0.02$pc) and 560\Msun~(for $R_{\mathrm {out}} = 0.04$pc), which are almost negligible compared to the total stellar mass of $10^4$\Msun~(and imply $\gtrsim 95$\% star formation efficiency).  This extreme choice of initial conditions is obviously highly contrived, but serves to set useful limits on the evolution of the system.

\begin{figure}
\centering
       \resizebox{\hsize}{!}{
       \includegraphics[angle=270]{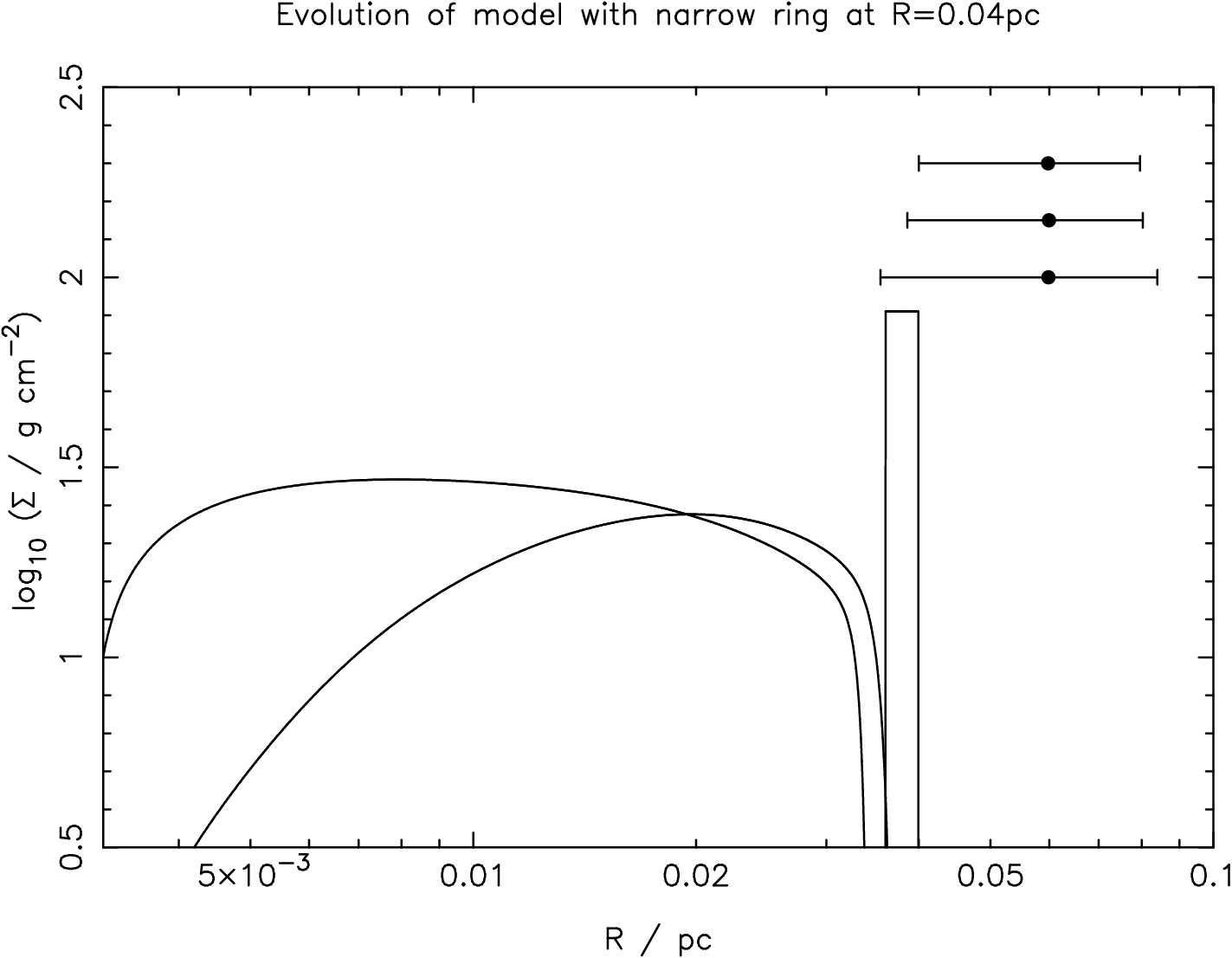}
       }
       
       \vspace*{6pt}
       
       \resizebox{\hsize}{!}{
       \includegraphics[angle=270]{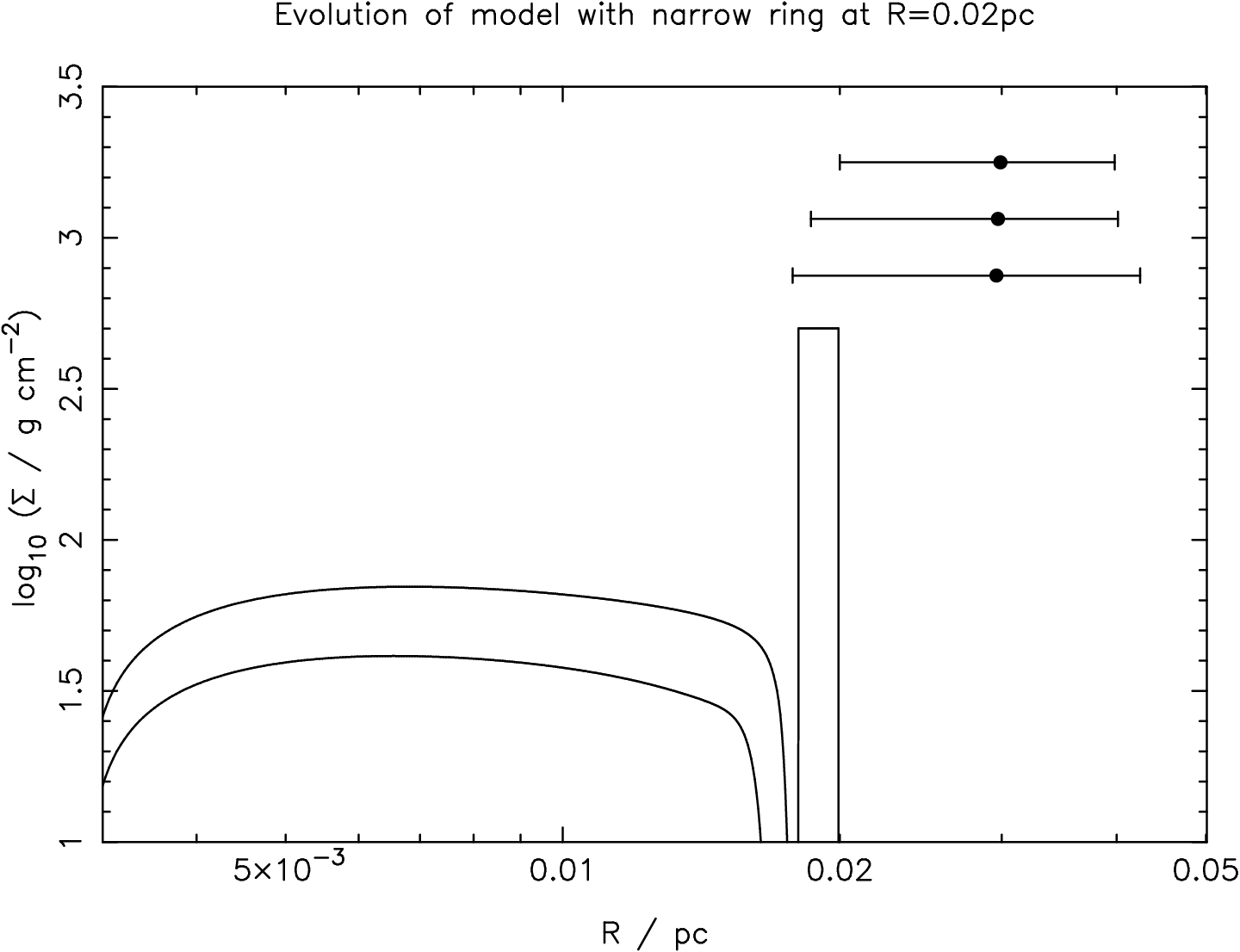}
       }
       \caption{Evolution of the disc-stars system where the `disc' initially consists of a narrow ring of gas just inside $R_{\mathrm {out}}$ (as described in Section \ref{sec:spikes}).   The upper panel shows the model with $R_{\mathrm {out}} = 0.04$pc; the lower panel the model with $R_{\mathrm {out}} = 0.02$pc.  As in Fig.\,\ref{fig:sigma}, the lines show the disc surface density, plotted at $t=0$, 5 \& 10Myr, and the filled circles and `error bars' mark the location of the stellar disc.  Even with this extreme choice of initial conditions a significant gas disc remains after 10Myr.}     
           \label{fig:spikes}
\end{figure}

The evolution of these models is shown in Fig.\,\ref{fig:spikes}.  Despite the implausible initial conditions we see that the system still evolves too slowly.  Even in these most optimistic models there is still a substantial gas disc present after 10Myr.  The total gas mass after 10Myr is 342\Msun~in the model with $R_{\mathrm {out}} = 0.04$pc and 129\Msun~in the model with $R_{\mathrm {out}} = 0.02$pc.  The accretion rates at the inner boundary are $6.9\times10^{-6}$\Msunyr~\& $1.5\times10^{-6}$\Msunyr respectively, and in both cases the surface density at 0.01pc exceeds 10g\,cm$^{-2}$.  We therefore see that even for an extreme choice of initial conditions, with an implausibly small residual gas disc and unreasonably efficient accretion, the combination of viscous accretion and tidal torques does not lead to accretion of the gas disc on a time-scale comparable to the age of the GC stellar discs.  This suggests that other mechanism(s) must be responsible for removing gas from the region close to the SMBH at the centre of the Galaxy.

\section{Discussion}\label{sec:discussion}
The most striking discrepancy between our model and the GC is in the SMBH accretion rate.  There is some evidence that the accretion rate on to Sgr A$^*$ may have been a significant fraction of the Eddington rate 5--10Myr ago \citep*[e.g.,][]{zubovas11}, but the present-day accretion rate is very low ($< 10^{-6}\dot{\mathrm M}_{\mathrm {Edd}}$, e.g., \citealt{bower03}).  We note, however, that this observed accretion luminosity tells us only about the physical conditions very close to the SMBH, and does not strongly constrain the presence or absence of an accretion disc at larger radii.  As discussed in Section \ref{sec:discs} (see also Fig.\,\ref{fig:mdot}), our models predict significantly sub-Eddington accretion 5--10Myr after the stars form, $\sim 10^{-4}$--$10^{-3}\dot{\mathrm M}_{\mathrm {Edd}}$.  A SMBH disc accreting at these low rates can in principle undergo radiatively inefficient accretion or outflow close to the Schwarzschild radius \citep[e.g.,][]{ny94,bb99}, and in this case the observed luminosity of Sgr A$^*$ is not necessarily inconsistent with the presence of an accretion disc at $\sim 0.01$pc radii.  Indeed, the favoured explanation for the low luminosity of Sgr A$^*$ is that the accretion flow is radiatively inefficient \citep*[e.g.,][]{yuan02}.  Detailed study of the accretion flow close to the SMBH (i.e., at radii $\lesssim 10^{-3}$pc) is beyond the scope of this paper, but we note that our models are not necessarily ruled out by observations of the Sgr A$^*$ accretion flow alone.

A more stringent limit on our models comes from study of the disc at larger radii.  An optically thick disc at radii $\sim 0.01$--0.1pc from Sgr A$^*$ would be easily detectable, both through its intrinsic emission and through reprocessing of the local radiation field.  At these radii the disc is rather cool, and the bulk of the accretion luminosity is radiated at mid-infrared wavelengths ($\sim 10$-100$\mu$m).  We can estimate the total flux emitted from the disc by using Equation \ref{eq:onezone} and assuming that each annulus in the disc radiates as a black-body.  Assuming a distance to Sgr A$^*$ of 8kpc, our range of models gives mid-IR flux densities that range from a few mJy to several Jy (with larger values corresponding to more massive and more compact discs).  Mid-IR observations of the inner $\sim0.1$pc are still rather scarce, but the recent high-resolution observations of \citet{schoedel11} show no evidence of this emission.  Moreover, the presence of an optically thick reprocessing disc is strongly ruled out by observations \citep[e.g.,][]{cuadra03,paumard04}, yet all of our models predict disc optical depths $\tau>10$ at these radii (even the extreme cases discussed in Section \ref{sec:spikes}).  Our models are therefore completely inconsistent with observations of the GC.

The basic reason for this discrepancy is easy to understand: at a radius $\sim0.1$pc, the characteristic viscous time-scale $R^2/\nu \sim 10^7$--$10^8$yr.  This is an order of magnitude larger than the age of the GC stellar discs, so it is obvious that viscous accretion alone cannot remove the residual gas disc.  We have shown that the combined effects of viscous accretion and dynamical interactions with the observed stellar disc are still insufficient, and simply do not remove angular momentum from the gas rapidly enough.  Our model is constructed in the context of the disc fragmentation model for GC star formation, but we note that our conclusion applies to any model which forms stars {\it in situ} with $<100$\% efficiency.  It is inconceivable that $\sim10^4$\Msun~of stars could have formed at their current location without creating a small residual gas disc \citep[cf.,][]{br08,hn09}, but we have shown that even $\sim100$\Msun~of gas cannot be accreted in a plausible time-scale.  The key question, therefore, is what happened to this `missing' gas?

One potential solution would be substantially enhanced angular momentum transport in the disc.  Our models with $\alpha = 1$ rule out any `local' transport mechanism (i.e., any mechanism where the accretion energy is dissipated locally).  Recent models suggest that strongly magnetised discs accrete at greatly enhanced rates \citep*{jl08,gaburov11}, but it is not yet clear if this mechanism is efficient enough to drive the accretion required here.  Alternatively, if the disc is self-gravitating then low-order spiral density waves can drive non-local transport \citep[e.g.,][]{bp99}.  This is, however, unlikely to be significant here.  The discs considered here are very thin (with $H/R \sim 10^{-3}$), and non-local effects in self-gravitating discs generally only become significant when the disc aspect ratio $H/R \gtrsim 0.1$ \citep[e.g.,][]{lr05,durisen07}.  Moreover, the radiative cooling time of a self-gravitating disc at the GC is very short \citep[e.g.,][]{nayak06,levin07}, so if the disc becomes self-gravitating it will fragment and form more stars, rather than accreting.  Self-gravity can in principle explain either the formation of stars or enhanced rates of accretion, but not both.

A more interesting solution is the idea that the residual disc could be removed via interaction with additional gas with a different angular momentum vector.  `Accretion events', such as the capture of a gas cloud invoked in Section \ref{sec:intro}, are expected to be relatively frequent, and there is no observational or theoretical reason to expect successive such events to have similar geometries \citep{kinney00,kendall03}.  This picture of chaotic accretion has previously been suggested as a mechanism for GC star formation \citep{kp06,kp07}, and carries the additional benefit of allowing accretion to proceed at close to the Eddington rate without significantly increasing the black hole spin \citep{klop05,king08}.  Capture of a gas cloud with a different angular momentum vector can lead to cancellation of  angular momentum and very rapid accretion \citep[e.g.,][]{hobbs11}, and this represents a plausible explanation for the missing gas disc around Sgr A$^*$.  Further work is needed to determine what, if any, additional observational signatures this model would predict.  However, chaotic accretion may explain both the presence of young stars and the absence of a gas disc at the GC, and provides a natural link between the observed star formation close to Sgr A$^*$ and the larger question of how SMBHs accrete.

If we accept that Sgr A$^*$ somehow accreted the `missing' gas disc, with mass $\sim 10^3$--$10^4$\Msun~\citep[e.g.,][]{hn09}, then the implications for the Milky Way are also interesting.  Since this gas should have been accreted on a time-scale much shorter than the age of the central starburst \citep[$\simeq 6$Myr, e.g.,][]{paumard06}, then the average accretion rate on to the SMBH must have been a significant fraction of the Eddington rate during this putative accretion episode.  As mentioned above, this is in fact consistent with recent observational \citep*{su10} and theoretical \citep{zubovas11} evidence that Sgr A$^*$ produced AGN-like feedback for $\sim0.1$--1Myr at approximately the same time as the stellar disc formed.  If this is indeed the case, the emergent chronology of the central parsec of the Milky Way is consistent with (1) deposition/capture of a large molecular cloud; (2) formation of a gas disc (or discs), the outer part of which rapidly fragmented into stars; (3) subsequent accretion of the interior remnant gas disc, resulting in the production of the $\gamma$-ray emitting kpc-scale Fermi lobes.  However, the detailed dynamics of these processes remain uncertain, and further work is needed if we are to understand this scenario more thoroughly.

\section{Summary}\label{sec:summary}
We have presented models of disc evolution following the formation of the young stellar disc at the Galactic Centre.  Our models include viscous accretion of gas and tidal torques between the young stars and the disc, as well as gravitational interactions between the stars.  We find that these processes do not remove the residual gas disc quickly enough, regardless of the choice of initial conditions and model parameters.  Instead a substantial gas disc always remains, interior to the stellar disc, for $>10$Myr after the stars form.  Such a disc is completely ruled out by observations, but the fate of this `missing' gas is unclear.  The most plausible mechanism for accreting the residual gas disc is direct cancellation of angular momentum following a subsequent accretion event, but further work is needed to understand the dynamics of this process in detail.  Our results again highlights the link between Galactic Centre star formation and SMBH feeding, and we suggest that chaotic accretion of gas clouds provides a plausible explanation for both the observed stellar disc and the accretion history of Sgr A$^*$.

\section*{Acknowledgements}
RDA acknowledges support from the Science \& Technology Facilities Council (STFC) through an Advanced Fellowship (ST/G00711X/1).  SLS thanks the University of Leicester for a SURE summer studentship.  Theoretical Astrophysics in Leicester is supported by an STFC Rolling Grant.  This research used the ALICE High Performance Computing Facility at the University of Leicester. Some resources on ALICE form part of the DiRAC Facility jointly funded by STFC and the Large Facilities Capital Fund of BIS.


\label{lastpage}

\end{document}